\documentclass[letterpaper]{article} 
\usepackage{aaai2026}  
\usepackage{times}  
\usepackage{helvet}  
\usepackage{courier}  
\usepackage[hyphens]{url}  
\usepackage{graphicx} 
\usepackage{natbib}  
\usepackage{caption} 
\usepackage{algorithm}
\usepackage{algorithmic}

\usepackage{subcaption}
\usepackage{pdfpages}
\usepackage{newfloat}
\usepackage{listings}
\DeclareCaptionStyle{ruled}{labelfont=normalfont,labelsep=colon,strut=off} 
\floatstyle{ruled}
\newfloat{listing}{tb}{lst}{}
\floatname{listing}{Listing}
\title{Cost-of-Ethics Crisis: Beliefs, Decisions, and Justifications in the Job Searches of Computer Science Students in Canada and the United States}
\author{
    Mohamed Abdalla\textsuperscript{\rm 1}, Sahar Abdalla\textsuperscript{\rm 2},
    Alicia Cappello\textsuperscript{\rm 3},
    Kyrie Dowling\textsuperscript{\rm 4}, \\
    Dana\'e Metaxa\textsuperscript{\rm 4},
    David G. Widder\textsuperscript{\rm 5},
    Catherine Stinson\textsuperscript{\rm 3}
}
\affiliations{
    \textsuperscript{\rm 1}University of Alberta, \textsuperscript{\rm 2}University of Toronto, \textsuperscript{\rm 3}Queen's University\\
    \textsuperscript{\rm 4}University of Pennsylvania, \textsuperscript{\rm 5}University of Texas at Austin
}

\usepackage{bibentry}

\begin{document}

\maketitle

\begin{abstract}
Workplace norms in computer science have received growing attention due to a  series of recent ethical scandals. One response has been a push to improve the ethics education provided to computer science students. Evidence for the effectiveness of ethics education remains mixed; some evidence suggests that norms are changing, others point to persistent gaps between stated values and practice remain. In this paper, we explore whether students, who have received some contemporary CS ethics education, are able to effectively apply ethical reasoning to their own decision-making in what is typically the first significant ethical decision of their careers: the job search.
Our study examines the ethical decision making of 129 computer science students and recent graduates during their job searches. We find that most students prioritize factors like compensation, location, and workplace culture over ethical and social issues. Even when expressing ethical concerns, respondents often justify taking actions contradicting their moral views through commonly-shared explanations such as desire to make money or the perceived inability to avoid unethical workplaces. This work sheds light on the possible disconnect between the ethics education CS students receive and the decisions they make during their job search. We offer insights for evolving curricula to better address practical ethical dilemmas, and encouragement to reproduce the study on a larger scale. 
\end{abstract}


\section{Introduction}
\begin{quote}
    \textit{``I literally will take an internship from Satan if it means I don't go homeless.'' -- survey respondent}
\end{quote}
The increasing digitization of nearly every aspect of modern life has brought the ethics of computing into the spotlight. 
High-profile scandals -- from the misuse of personal data at scale by platforms \cite{wylie2019mindf} to the deployment of biased facial recognition systems \cite{cook2025performance} -- have made clear that current norms are insufficient to ensure that the computer science (CS) workforce acts responsibly and produces positive societal outcomes \cite{kumar2024computer}. 
While multiple professional codes exist (e.g., ACM's Code of Ethics and Professional Conduct \cite{anderson1992acm,gotterbarn2018acm}, ABET accreditation standards \cite{abet2004criteria}, and others \cite{acscode,bcscode,nzcode}), many critics believe they are too abstract to be able to drive real change in decision making \cite{mcnamara2018does,gogoll2021ethics}.
Researchers have similarly enumerated over 50 ethical principles specific to AI \cite{munn2023uselessness}, but criticisms range from being too abstract \cite{zhou2023ai} to being ``\textit{meaningless, isolated, toothless and adher(ing) to industry's agenda}'' \cite{munn2023uselessness}.

In response, there has been a growing push to improve both the quantity and the quality of ethics education provided to CS students \cite{fiesler2020we}, premised on the assumption that better ethical training will translate into more ethical behavior, decision making, and improved societal outcomes. To address these issues, educators have been actively trying to embed ethics into more technical CS courses, as well as updating their CS curricula to include more ethics courses \cite{skirpan2018ethics,horton2023more,stavrakakis2021teaching,horton2022embedding,brown2024teaching}. We use the term ``ethics'' broadly, following past work \cite{stavrakakis2021teaching,sarder2022entering}, to encompass computing ethics, information ethics and related fields as well as concepts like social responsibility and justice.

Yet evidence that this educational push is working remains mixed. On one hand, some CS practitioners have taken ethical concerns seriously: most AI developers, when asked, now affirm that ethics is central to their work \cite{vakkuri2019ethically}, and students have confronted companies from Facebook to Lockheed Martin and Chevron in interviews and career fairs \cite{Rodriguez2019Struggle,flores2024lockheed} with more difficult, ethics-oriented questions. On the other hand, studies consistently report a gap between stated values and practice. When studying the priorities of engineering students at 4 different colleges, \citet{cech2014culture} noted that ``\textit{public welfare concerns are not highly valued in students’ professional identities as engineers}.'' Students self-report that they often found lessons from ethics courses difficult to apply to their own work \cite{abraham1997experiences}, a finding which has been corroborated by others \cite{Reidy2017shock,oliver2021undergraduate}. US and Australian data science curricula often fall short of equipping students for real ethical decision making \cite{oliver2021undergraduate} and tend to emphasize rule-following over broader societal responsibility \cite{gorur2020computer,abdurrahman2023amazon,schiff2021linking,papak2018examining}.  A follow-up study by \citet{vakkuri2020just} of AI developers at startups found that, despite agreeing that AI ethics was theoretically important, it was not relevant to them during early-stage development.

We conceptualize the job-seeking process as the first real-world site where ethical commitments are enacted, negotiated, or set aside -- making it a uniquely revealing lens through which to study how ethics education is translated into actual behaviour. Previous studies of the effects of ethics education have several limitations. We believe asking workers whether they care about ethics is insufficient, as it fails to capture the complexity of contextual decision making. An individual's commitment to ethics may be relative to other competing priorities. Furthermore, self-reported attitudes may not accurately reflect actual behaviour, making a comparison between stated beliefs and actions a more informative approach. Lastly, the specific decision-making processes and rationales used during actual ethical conflicts, as opposed to answers to hypothetical questions, can provide valuable insights into the effectiveness of current ethics education, and may inform future curriculum development. To that end, we pose the following research questions:

\begin{itemize}
    \item \textbf{RQ1}: What is the relative importance of ethical considerations compared to other factors (e.g., compensation, work-life balance, etc.) during the employment search?
    \item \textbf{RQ2}: Which specific ethical issues are most concerning to students?
    \item \textbf{RQ3}: How do students rationalize working for companies involved in activities they find ethically concerning?
\end{itemize}

To answer these questions, we surveyed CS students in Canada and the United States who were actively seeking employment. Participants ranked job considerations, identified companies they had applied to, identified their most pressing ethical concerns, and explained their reasoning when those companies' records conflicted with their stated values. Our analytic interest is not in passing judgment on these choices, but in examining the structure of the justifications students offer -- and what those justifications reveal about the reach and limits of current ethics education. This work speaks directly to CS ethicists' interest in the human and organizational factors that shape whether ethical commitments in computing translate into practice. The students in our study are tomorrow's CS workforce -- the values they bring to their jobs will influence the technologies they build and the broader culture of the industry \cite{shen2021value}.

Our paper makes the following contributions: For a small sample of job-seeking CS students, we:
\begin{enumerate}
    \item Quantify the perceived importance of ethical considerations relative to other factors in employment decisions, revealing their comparative lack of prioritization.
    \item Characterize the decision-making processes students employ to rationalize applying to companies engaged in activities they deem ethically concerning.
    \item Outline the implications of our results for students, educators, and industry.
\end{enumerate}

\section{Background}
\subsection{Computer Science Ethics Education and Critiques}
Researchers have long been interested in the effective teaching and assessment of ethics across various academic disciplines \cite{hedayati2020understanding}. The goal of ethics education in CS, per the ACM/IEEE-CS/AAAI CS2023 curricular guidelines, is to develop students' understanding how society influences CS, how CS influences society, and their role in this relationship \cite{kumar2024computer}. Traditionally, in CS, ethics was taught as a standalone course, yet a growing number of scholars argue that teaching ethics separate from technical material serves to further cement the view that ethics is not an integral component of CS \cite{fiesler2018our,grosz2019embedded}. As a result, we have seen an increasing number of technical courses incorporating ethics modules \cite{brown2024teaching,stavrakakis2021teaching,jarzemsky2023applies,oleson2025design}, though a large number (38\%) of institutions still teach computer ethics as a standalone course \cite{stavrakakis2021teaching}. Polling European faculty in CS and related disciplines, \citet{stavrakakis2021teaching} found that 36\% of respondents believed their institutions did not teach `enough' ethics as part of the curriculum (and 63\% of respondents affirmed that teaching ethics was `Very Important' or `Important'). 

Surveys of curricula show that instructors teaching ethics use a variety of pedagogical tools (e.g., lectures, classroom discussions, and essay assignments) \cite{pant2024teaching,rivera2024teaching}, with most of the examined syllabi focusing on teaching students how to recognize ethical issues, assess and evaluate these issues, and make reasoned arguments \cite{fiesler2020we}. To evaluate the effect of these educational efforts on students, most courses rely on proxy measures such as students' self-reported satisfaction and awareness of ethical dilemmas collected through surveys \cite{horton2022embedding,horton2023more,skirpan2018ethics,brown2024teaching} or student feedback and writing \cite{skirpan2018ethics}. 

As an example, \citet{horton2022embedding} created a survey, Ethics Attitudes and Self-Efficacy (EASE), to assess students' self-reported interest in ethics and technology, as well as their self-reported self-efficacy in dealing with ethical issues. Studying first-year CS students, they found that students improved from their baseline when an ethics module was incorporated in their technical course. In a follow-up study, \citet{horton2023more} compared the effects of embedding one versus multiple ethics modules across multiple semesters and found that: 1) the benefits of the \textit{first} embedded ethics module are robust (i.e. reproducible), but 2) having multiple embedded ethics modules in a semester did not increase self-reported efficacy, and 3) EASE scores dropped between semesters but could be improved, back to the original level, by embedded ethics modules in subsequent semesters. Similarly, following the pilot of the Embedded EthiCS program at Harvard, \citet{grosz2019embedded}, measured that, across all the courses, 85\% of students agreed with the statement ``\textit{The ethics guest lecture increased my interest in learning about the moral issues we discussed}'' and ``\textit{I would be interested in learning
more about ethics in future computer science courses}.''

The above efforts assume that more ethics education will produce practitioners who are more ethically knowledgeable and, consequently, more ethical actors. However, this assumption is not universally shared by educators or students \cite{widder2023s,tenbrunsel2004ethical,rivera2024teaching}, as students may feel unsure about how to apply the lessons and discussions to their future work \cite{rivera2024teaching}, while researchers argue that unethical decision making is not due to a lack of formal education in ethics, but rather to `psychological processes' which enable the unrecognized violation of ethical principles held by the decision makers \cite{tenbrunsel2004ethical}. Other work suggests that the lack of ethical decision making is partly due to the process of obtaining industry jobs, internships, and lab placements. When these highly desired positions only value technical prowess, it sends a message to students that critical ethical evaluation is not necessary or beneficial \cite{darling2024not,kirdani2024taught,sarder2022entering}.

Evaluating the likelihood that students will use ethics knowledge that was taught to them (i.e., put their education into practice) is difficult. As noted previously, most studies rely on surveying students' self-reported interest in ethics as a proxy for the likelihood of applying what they've learned. Only a few studies have conducted evaluations of, or observed, students applying or demonstrating their ethics knowledge in real-world scenarios \cite{brown2024teaching}.

\subsection{Decision-making processes of computer science students and workers}
Behavioral ethics is “\textit{the study of systematic and predictable ways in which individuals make ethical decisions and judge the ethical decisions of others when these decisions are at odds with intuition and the benefits of the broader society}” \cite{bazerman2012behavioral}. Behavioral ethics is interested in the processes and rationalization behind ethical decisions \cite{folger2012deonance}.

Researchers have described the many ways decision makers make unethical choices while believing they are upholding their moral principles -- a process termed `ethical fading' by Tenbrunsel and Messick \cite{tenbrunsel2004ethical}. For example, one may use `language euphemisms' to re-interpret their specific context in a way which changes its moral framing or intensity. They also highlight how repeated exposure to an ethical decision may result in `psychic numbing' which, when combined with the natural minimization of moral weight to small changes, can result in many small, repeated unethical decisions being made (termed `slippery slope'). Furthermore, ethical assessments may be misinformed, erroneous, or affected by self-interest \cite{tenbrunsel2004ethical}.

Focused specifically on computer scientists, Hedayati-Mehdiabadi studied the decision-making processes of CS students enrolled in a computing ethics course at a small Midwestern university in the United States \cite{hedayati2022computer}.  The students were asked to imagine different hypothetical scenarios (e.g., selecting different privacy levels for an application or delivering potentially buggy software to meet deadlines). Depending on the perceived moral intensity of the decision, there was often a single response favored by the super-majority of students (e.g., selecting the highest privacy setting as default). Various factors affected the rationalization of students such as perceived ethical significance, personal history, and assumptions about the end-user. Hedayati-Mehdiabadi noted that, despite being exposed to various ethical theories, few students referenced (or seemingly applied) these theories in their decision making \cite{hedayati2022computer}. 

Past research has also demonstrated that the application of ethics by practitioners is not straight-forward. For example, \citet{widder2023dislocated} demonstrated how software engineers see the ethical implications of what they build as too far removed from their work (occurring many steps beyond their stage in the "AI supply chain"), thus allowing the modular nature of the AI supply chain to enable ethical distancing. Even if ethical concerns were identified by the engineers in a nearby stage of the ``AI supply chain,'' the engineers often lacked the power to address these concerns directly \cite{widder2023s}. As \citet{widder2023s} showed, these concerns typically related to the use of the technology under development, where the engineers had limited influence, rather than the design of the technology. Interviewing students enrolled at the University of Colorado Boulder, \citet{sarder2022entering} noted further that it was difficult to get students to value ethics due to their perception that employers did not value it.

\section{Methods}
\subsection{Survey}
This study was approved by the research ethics boards of Queen's University, University of Alberta, and University of Pennsylvania and the survey was hosted using Queen's Qualtrics platform. The full survey can be found in the Appendix. Two nearly identical versions were used, one for Canadian students and one for American students, with only minor adjustments made between them (e.g., aligning with each country's specific census categories for personal identification). The survey is composed of three sections. The first section deals with demographic information, companies applied to, and ranking a predetermined list of priorities. Individuals responding to our survey (henceforth: respondents) are also able to add a priority not found on the predetermined list. The second section of the survey asks respondents if they have any concerns about their labour being used: 1) for military purposes, 2) by a company engaged in exploitative working conditions (e.g., forced labour or wage theft), 3) to negatively impact the environment, 4) to invade the privacy of others, or 5) to negatively impact the mental health of others -- a selection of ethical issues  informed by past work \cite{widder2023open}. The last section of the survey is dynamic, presenting each respondent with different questions by comparing their responses about companies applied to (first section) with their stated levels of concern (second section). If a potential conflict is  detected -- for example, when a respondent expresses strong concern about their labour being used for military purposes while also indicating an interest in working for companies such as Alphabet or Amazon, which have well-documented military contracts \cite{gonzalez2024big} -- the respondent was asked to elaborate on their awareness of this tension and their thoughts regarding it.

To mitigate participation bias (also known as non-response bias) \cite{sedgwick2014non}, survey participants were told that the purpose of the survey was to assist CS educators in understanding student preferences and views related to the job search. To avoid order effects bias \cite{perreault1975controlling}, the questions were structured such that more innocuous questions (e.g., `What companies have you applied to?') were asked before questions more likely to elicit a defensive response (e.g., `What are your views of your labour being used for military purposes?'). To minimize framing effects, we iteratively worked to improve the phrasing by beta-testing our survey among colleagues and personal connections several times. More details about the survey methodology can be found in Appendix \ref{app:surveymethods}.

We defined the target demographic as any CS student currently attending a Canadian or American post-secondary institution seeking opportunities post-graduation (e.g., final year undergraduate students), or recent CS graduates who had just completed a job search. We sent emails to university departments and professors teaching 4th year courses. We also posted to university subreddits (where permitted by the subreddit rules/moderators).

Despite our best efforts, our results are subject to various biases common in survey methodology. A potential non-response bias exists, as participants who complete surveys without participatory incentives are often more socially engaged \cite{porter2004understanding}. Furthermore, sampling bias from our target population limits the generalizability of our findings. Additionally, while our sample size is in line with past studies \cite{hedayati2020understanding,donaldson2012organisational,widder2023s}, it is too small to be representative of our target demographic. Lastly, for our subgroup analysis, we combined some categories due to practical sample-size considerations and to enable cross-comparison between American and Canadian respondents. \textbf{Subgroup results should be read as suggestive and exploratory rather than as robust, generalizable claims about these groups}.

\subsection{Qualitative Analysis}
To analyze the free-text responses, two authors performed an open qualitative card sort \cite{zimmermann2016card,widder2023s,widder2019conceptual} to assign each response to cohesive categories. Categories (e.g., economic pressure, role/proximity distancing, prestige, etc.) were developed inductively from the responses themselves rather than applied from a pre-existing schema, though annotators were informed by previous work on behavioral analysis. Two coders independently coded a subset of responses then coded the remaining responses using this shared scheme. Disagreements were negotiated through discussion until consensus and categories were adjusted as required. The resulting analysis was then reviewed by other authors.

\subsection{Quantitative Analysis}
We quantitatively analyze both complete and partial responses -- partial responses were included in the analysis only if the respondent completed up to and including the second section (regardless of the completion of the third section). Note, per the REB-approved protocol, respondents were not required to complete all/any questions. As such, there are some missing responses. For transparency, we report ratios and percentages in the analysis, as well as denominators. Full results can be found in the supplemental material.

Our first round of quantitative analysis involves providing descriptive ratios for a subset of comparisons that have large apparent differences (i.e.,  one proportion or average rank being at least roughly double the other or a large visible difference). Next we evaluate the statistical significance for the comparisons that met this threshold. When comparing the ranked annotations (for the considerations) we used the Mann–Whitney U test (i.e., Wilcoxon rank-sum test) -- appropriate for comparing rankings. To compare between concern levels we use Fisher’s exact test -- a non-parametric test for binary categorical variables. To reduce the risk of post hoc significance testing (i.e., p-hacking), we do not evaluate statistical significance for any additional comparisons, rather only for those initially identified as having large apparent differences. For these comparisons, we report the corresponding p-values.

\section{Results}
All analyses below were conducted using the 129 complete and partial responses. In our analysis, we report ratios and percentages, as well as denominators. Full results can be found in the supplemental material.

Most respondents (70\%) were 4th year undergrads or had recently graduated (n=90). All results presented below (e.g., other sub-categories) consider only 4th year undergrads or recently graduated respondents, unless explicitly stated. For respondents in their 4th year or those who recently graduated, we had the following splits\footnote{For many categories, we do not report counts to maintain the privacy of respondents in line with the approved REB applications.}: Canadian at 61 vs US at 29; Male at 55, non-Male\footnote{We group all other genders under this label together as gender-minoritized/gender-marginalized in CS.} at 35; White at 40 and Asian\footnote{Small cells are not reported due to privacy concerns as specified in the REB submission.}  at 40. We note that all respondents were from accredited schools which taught AI ethics in some capacity (e.g., standalone courses, incorporated modules, etc).

\subsection{Quantitative Results}
For each question, we start by presenting the responses of all 4th year undergrads and recently graduated students, followed by sub-group analyses.

\subsubsection{Applying to Industry}
The majority of respondents (65\%) applied to at least one Big Tech company (pre-specified as Alphabet, Amazon, Apple, Meta, Microsoft\footnote{There is no consensus on which companies should be included when referring to so-called ``Big Tech'' \cite{abdalla2021grey,abdalla2023elephant,ahmed2020democratization,widder2023open,birhane2022values}. We chose a restrictive set of companies due to the branching required to make this survey dynamic.}). Appendix Table \ref{tab:application_percentage2} shows the full results. We observe no statistical difference in the application rates of US applicants vs Canadians (0.76 vs 0.67, $p<0.47$) or between Male and non-male applicants (0.73 vs 0.66, $p<0.49$) in applying to Big Tech. Asian respondents applied at significantly higher rates compared to White respondents (0.82 vs 0.58, $p<0.01$). Combined, University of Waterloo and University of Toronto (UofT) students applied at significantly higher rates compared to Queen's University students (0.83 vs 0.40, $p<0.04$)\footnote{These three schools had the largest number of respondents. UofT and Waterloo are actively recruited from by Big Tech and have a more competitive culture than Queen's University.}. We also observed differences between company application rates (e.g., across all respondents, Meta and Apple had fewer applicants) but trends were not consistent. 

\begin{table}
\centering
\setlength{\intextsep}{1pt}
\scalebox{0.8}{
\begin{tabular}{l|cccccccccccc}
\textbf{Respondent Type} & \textbf{All} \\ \hline
\textbf{Remuneration} & 2.31 \\
\textbf{Location}& 3.47  \\
\textbf{Work Life Balance} & 3.38 \\
\textbf{Workplace Culture} & 4.09  \\
\textbf{Ethical Concerns}  & 5.01 \\
\textbf{Prof. Dev. Opps}   & 3.26  \\
\textbf{Other}   & 6.49 
\end{tabular}
}
\caption{\label{tab:consideration_avgs_first} Average rank of job considerations for all respondents.}
\end{table}

\subsubsection{Job Considerations}
Table \ref{tab:consideration_avgs_first} presents the average ranking for all the predetermined considerations: Remuneration, Location, Work-life Balance, Workplace Culture, Ethical Concerns, and Professional Development Opportunities. Across all applicants, we observed that `Remuneration' ranked the highest (with an average rank of 2.31 out of 7) and `Ethical Concerns' ranked last (with an average rank of 5.01). We also observed differences in rankings across sub-groups, Appendix Table \ref{tab:consideration_avgs2}. For example, Male respondents ranked `Location' significantly lower (3.76 vs 2.97, $p<0.05$). Respondents from Queen's ranked `Work-life Balance' much higher than respondents from UofT or Waterloo (2.73 vs 3.81, $p<0.06$), though the difference was not significant. Respondents also had the option to specify their own considerations. Self-defined considerations included perceived company reputation, ability to leverage position for career progression, and enjoyment/mental stimulation of work.

\begin{figure}[h!]
    \vspace{0pt}
    \centering
    \includegraphics[width=\linewidth]{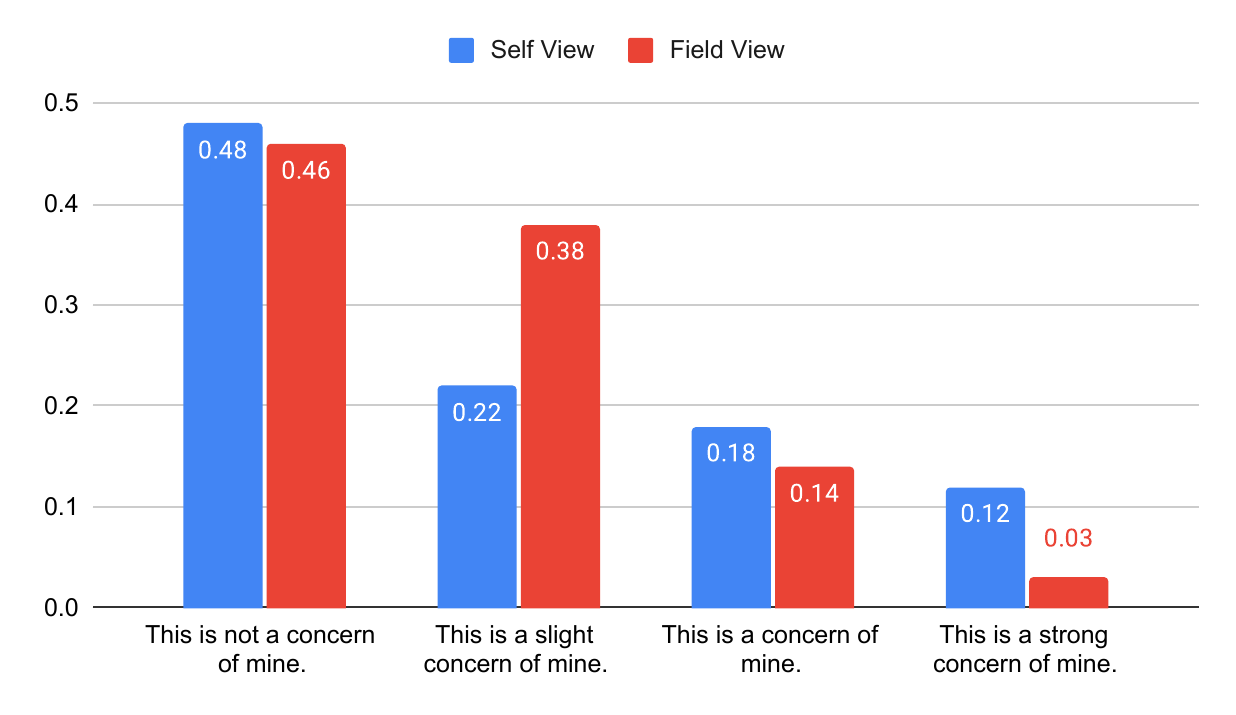}
    \caption{The decimal percentage of respondents who agreed with specific statements regarding the use of their labor for military purposes, categorized by Self-View (individual opinion) and Field View (perceived disciplinary perspective).}
    \label{fig:military_self_field}
\end{figure}
\subsubsection{Labour for Military Use}
Figure \ref{fig:military_self_field} illustrates the distribution of views held by respondents regarding their labour being used for military purposes (Self View) and what they believed was the view held by their colleagues (Field View). Most respondents (48\%) had no concern with their labour being used for military purposes, and this was close to the percent of people who estimated that the field shared the same view. Conversely, while 30\% of respondents indicated that use of their labour for military purposes was a concern or a strong concern, respondents estimated that only 17\% of the field shared this view $p<0.04$) indicating that respondents perceive the field caring significantly less than the data suggests.  

There is a large variation between gender sub-groups for this concern (as shown in Appendix Table \ref{tab:percentage_military_views_nopair}). Male respondents were significantly more likely to have no concern regarding their labour being used for military purposes compared to non-Male respondents (0.58 vs 0.30, $p<0.02$).

\begin{figure}[h!]
    \vspace{0pt}
    \centering
    \includegraphics[width=\linewidth]{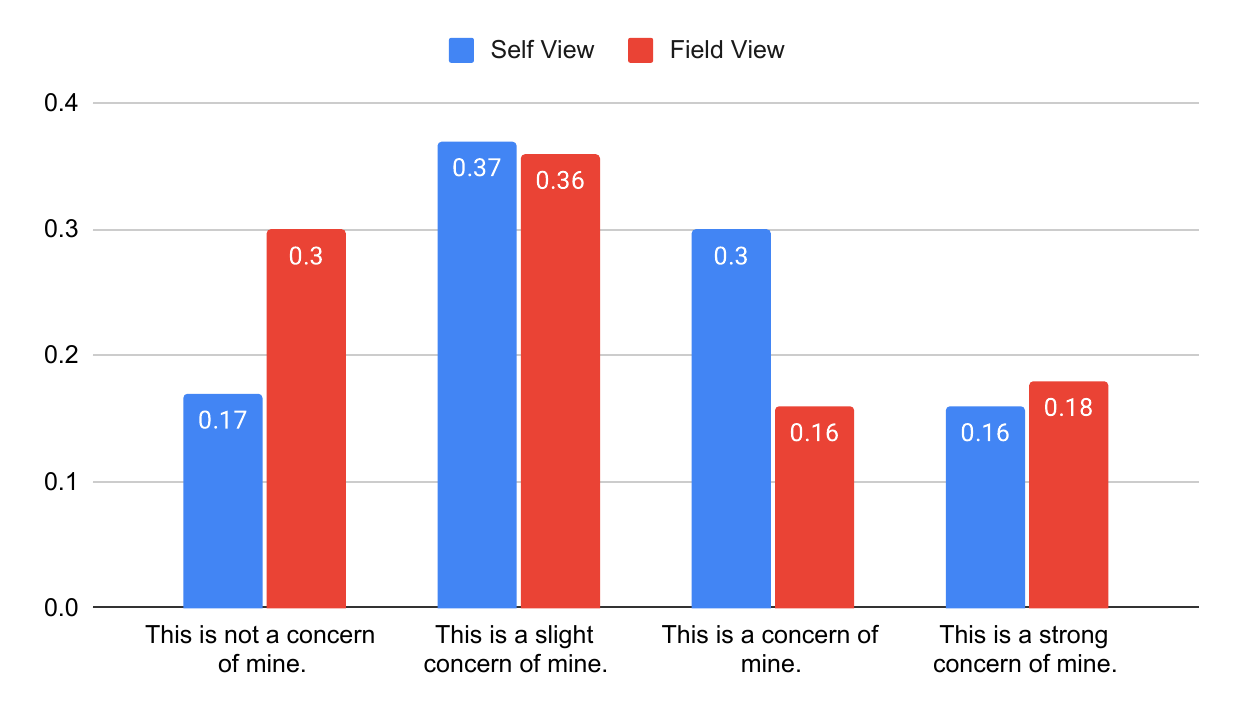}
    \caption{The decimal percentage of respondents who agreed with specific statements regarding the use of their labor by a company known to engage in exploitative working conditions, categorized by Self View and Field View.}
    \label{fig:exploit_self_field}
\end{figure}

\subsubsection{Labour for a company known to engage in exploitative working conditions}

Figure \ref{fig:exploit_self_field} illustrates the distribution of views held by respondents regarding their labour being used by a company known to engage in exploitative working conditions. The difference between the perceived lack of concern for this issue by the field (i.e., Field view compared to the respondents' self views) the difference was not statistically significant (0.30 vs 0.17, $p<0.09$). Likewise, proportion of male respondents with no concern or strongly concerned were not statistically significant (0.20 vs 0.11, $p<0.35$) and (0.12 vs 0.22, $p<0.35$) respectively, Appendix Table \ref{tab:percentage_exploit_views_nopair}. Likewise there was no statistically significant difference in the number of US or Canadian respondents  who reported strong concerns (0.23 vs 0.12, $p<0.11$).

\begin{figure}[h!]
    \vspace{0pt}
    \centering
    \includegraphics[width=\linewidth]{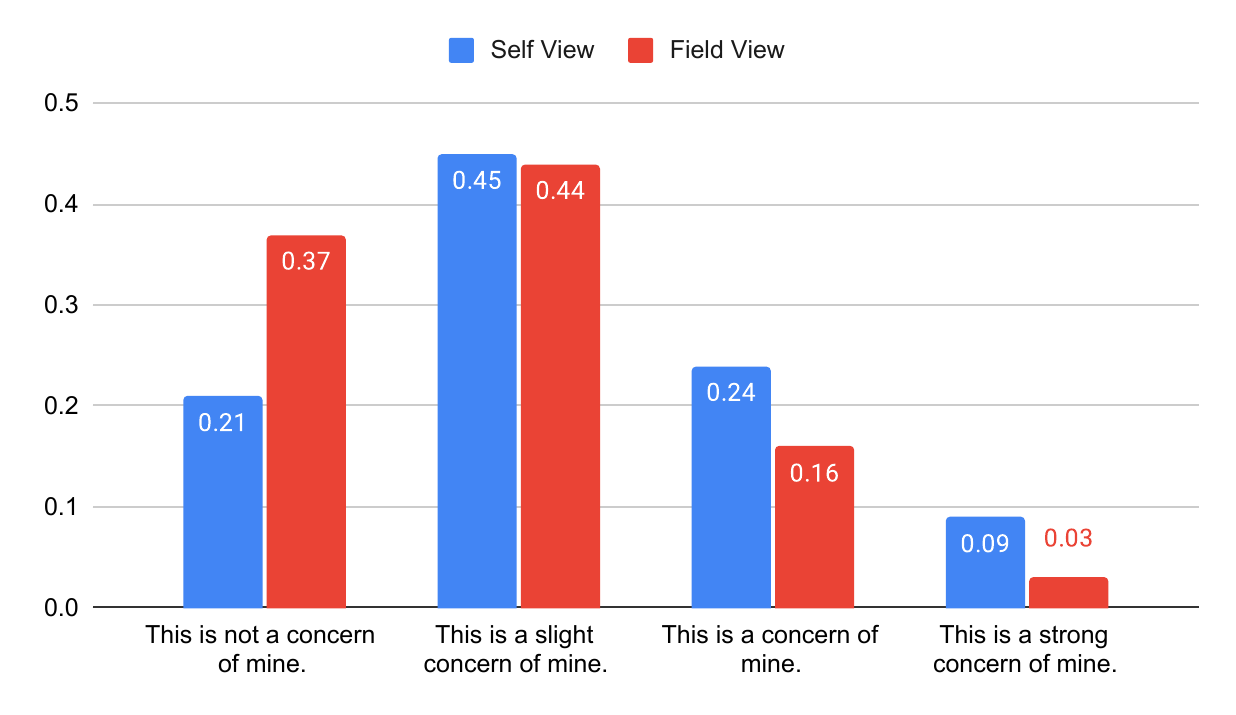}
    \caption{The decimal percentage of respondents who agreed with specific statements regarding the use of their labour by a company known to negatively impact the environment, categorized by Self View and Field View.}
    \label{fig:environment_self_field}
\end{figure}

\subsubsection{Labour for a company known to be negatively impacting the environment}

Figure \ref{fig:environment_self_field} illustrates the distribution of views held by respondents regarding their labour being used by a company that is negatively impacting the environment. The number of respondents with no concern was significantly lower than the proportion expected to have no concern (0.21 vs 0.37, $p<0.05$). The difference between the proportion of respondents expressing strong concern was not statistically significant (0.09 vs 0.03, $p<0.31$). 

Unlike previous comparisons, we did not observe any significant differences between sub-groups which has large numeric differences such as Canadians being more strongly concerned (0.12 vs 0.04, $p<0.41$), Appendix Table \ref{tab:enviorment_views}.

\begin{figure}[h!]
    \vspace{0pt}
    \centering
    \includegraphics[width=\linewidth]{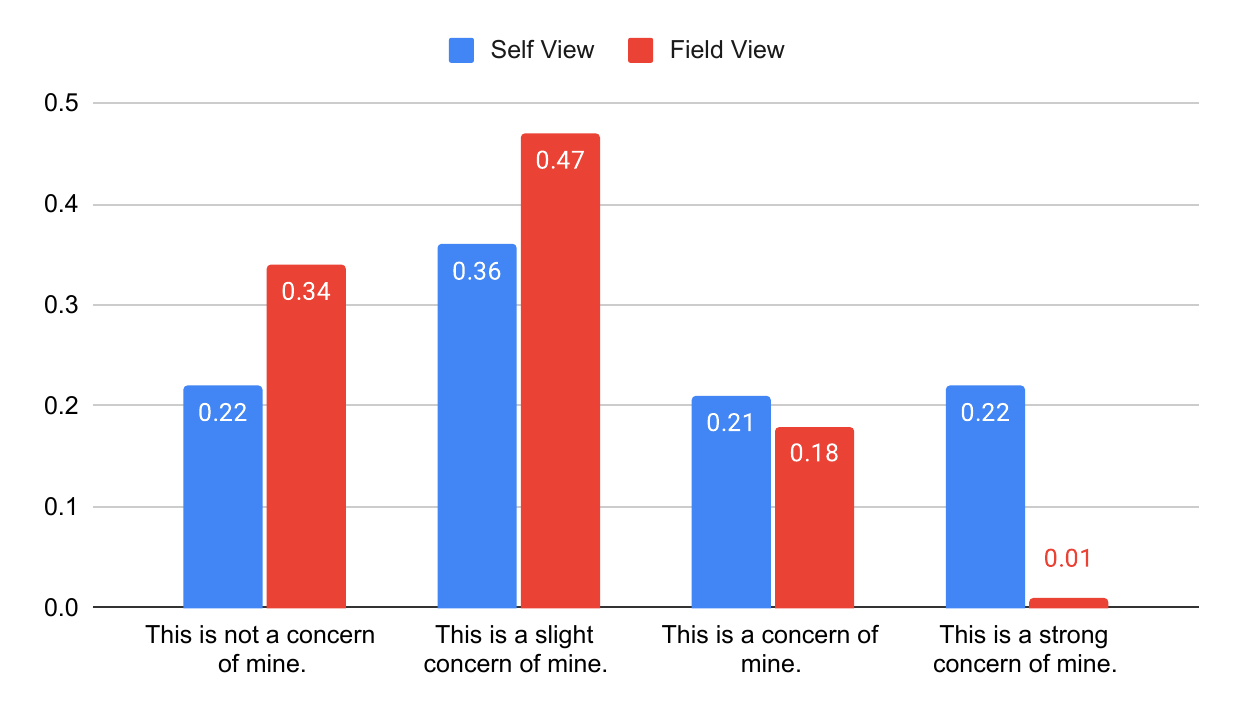}
    \caption{The decimal percentage of respondents who agreed with specific statements regarding the use of their labor by a company known to be invading the privacy of others, categorized by Self View and Field View.}
    \label{fig:privacy_self_field}
\end{figure}

\subsubsection{Labour for a company known to be violating privacy}

Figure \ref{fig:privacy_self_field} illustrates the distribution of views held by respondents regarding their labour being used by a company known to be violating the privacy of others, as well the field's perceived view. The proportion of respondents feeling strong concern vs the perceived proportion was statistically significant (0.22 vs 0.01, $p<0.01$), \ref{tab:privacy_views}.

\begin{figure}[h!]
    \vspace{0pt}
    \centering
    \includegraphics[width=\linewidth]{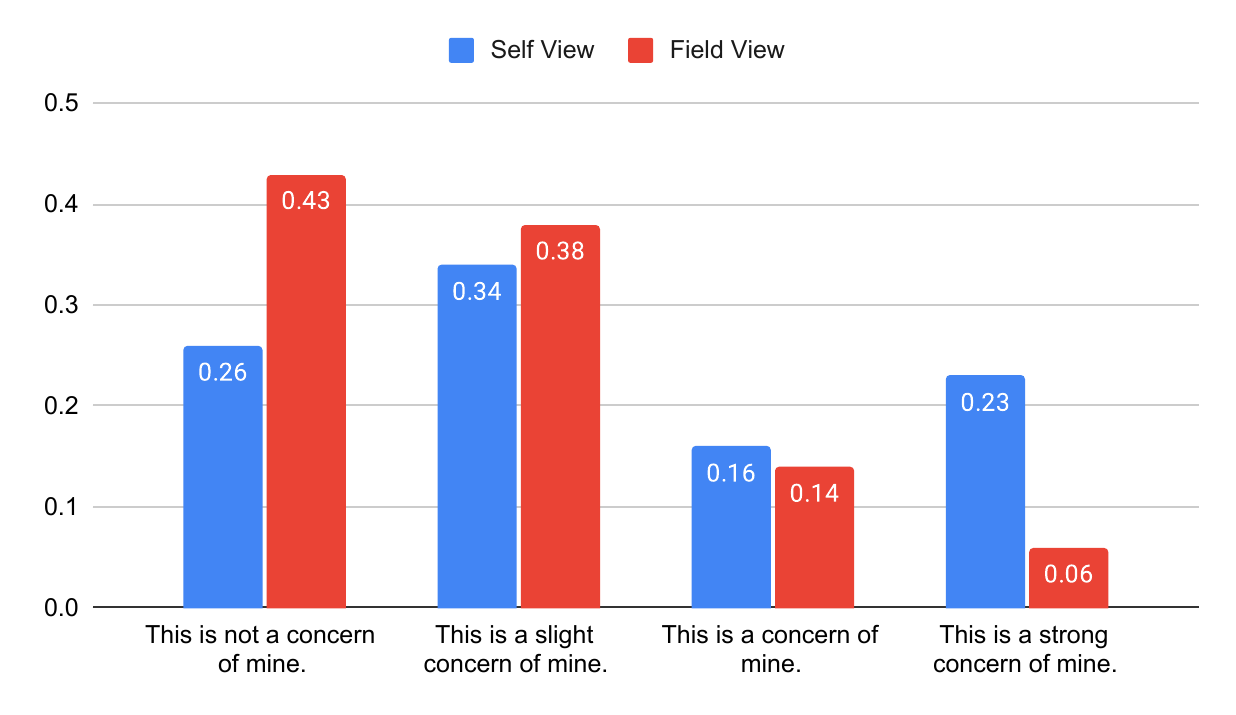}
    \caption{The decimal percentage of respondents who agreed with specific statements regarding the use of their labour by a company known to negatively impact the mental health of others, categorized by Self View and Field View.}
    \label{fig:mentalhealth_self_field}
\end{figure}

\subsubsection{Labour for a company known to be impacting the mental health of others}

Figure \ref{fig:mentalhealth_self_field} illustrates the distribution of views held by respondents regarding their labour being used by a company known to negatively impact the mental health of others, as well as what they believe best represented the field's view. In general, respondents felt the field's perceived views on this issue were significantly different from their own. The proportion of responders expressing having no concern was almost half the proportion perceived to have no concern (0.26 vs 0.43, $p<0.04$). Conversely, the proportion of respondents who expressed strong concern was more than triple the expected proportion (0.23 vs 0.06, $p<0.01$).

Appendix Table \ref{tab:mental_views} presents the results per sub-groups though we did not find any large differences between the sub-groups.

\subsubsection{Awareness of negative impacts}

For respondents who indicated concerns regarding working for a company known to negatively impact any of the associated concerns, but who had applied to a company with some connection to these negative effects or harmful practices, we presented examples of such behaviour and asked if they were aware of these instances. Awareness of examples varied greatly between respondents and issues with some level of awareness varying between 50\% and 83\%, Appendix Table \ref{tab:app_knowledge}.

\subsection{Qualitative Results}
Throughout the survey, respondents were prompted to share their thoughts regarding their selections. In this section we present analyses of the responses shared with us with some direct quotes to assess the ethical argumentation provided in justifying decisions.

\subsubsection{Justification for selected self views}
Respondents provided a large variety of explanations of their selected concern levels, some with multiple explanations in a single response. Most explanations were justifications of why their level of concern was not higher (despite neutral framing of the question). The most common responses (18\% of respondents) were classified as ``moral disengagement'' -- a process where to reduce feelings of guilt and to maintain a positive moral self-image, individuals disassociate their moral standards from their actions. This can be done in a variety of ways such as moral justification, where individuals re-frame their actions as moral, by displacing their responsibility, or by minimizing harm \cite{bandura1990selective}. Minimization of harm was particularly common on issues like privacy (``\textit{I feel protected knowing that they are doing this for millions upon millions of people -- what's the worst they could do to me? I don't have anything to hide}''). Quotes like this reflect a lack of understanding of the harms of surveillance \cite{solove2011nothing}.

Economic concerns (and the difficult job market) were a commonly provided explanation. The high level of economic anxiety experienced by Canadian students is reflected in statements like: ``\textit{I would be so grateful to have any type of computing science job at all}'', ``\textit{Given the state of the current economy with off-shoring and outright replacement of North American workers, I can't really restrict my scope}'', and ``\textit{I [literally] will take internships from Satan if it  means I don't go homeless. [It's] bad enough for CS grads with job listings plummeting by 35\%}''. Such concerns being top of mind is unsurprising given the recent economic downturn, where Canadian youth (aged 15-24) are experiencing a record high unemployment  rate of 14.5\% \cite{govcanyouth}. Canadian tech corporations offshoring jobs previously held by Canadians likely contributes to this sentiment among recent graduates \cite{Palermo2025unifor,hokiro2023telus,harris2017cibc,wilson2025hrd}. American respondents reported similar sentiments with one expressing, ``\textit{I have had to expand where I apply to, even if I don't like their practices, in order to increase my chances of receiving a job offer},'' and another stating, ``\textit{[anyone would] so long as the pay was good and it was a high-status job}'', reflecting the economic downturn facing the US tech sector \cite{ito2024recession}. Given that many students choose to study CS in pursuit of monetary compensation or financial stability \cite{downey2009mis}, it may be that compensation would have been a major motivation regardless of the status of the economy. Longitudinal analysis of responses is required to confirm this.
Several respondents explained that their responses heavily depended on the proximity of their personal work to the particular issue at hand (``\textit{It would depend on the specific project.}''). 

Other respondents were in complete moral alignment with tech corporations. This viewpoint tended to be more commonly and strongly held among US respondents. Students stated, ``\textit{It would be an honor to serve my country through my work},'' and regarding military applications that it ``\textit{would feel patriotic, satisfying, and fulfilling}''. Such views are not surprising given the environment in which many respondents live (e.g., Hollywood movies heavily lionizing the military \cite{weikle2020hollywood}, military flyovers at sporting events \cite{elbaba2022mlb}, etc.). A subset of the students in moral alignment also demonstrated an interesting view on invasions of privacy: ``\textit{Spying on others is a patriotic duty.}''  ``\textit{I'm happy if they do it to foreigners},'' but ``\textit{spying on Americans is not acceptable}.'' Some noted that they ``\textit{hate companies collecting [their] data}.'' The implication of these statements is that the ethicality of the action depends on \textit{who} is directly impacted. 

\subsubsection{How does this knowledge change your decision to apply to [Company]?}
If the respondent indicated that knowledge presented to them would change their decision to work for/apply to a company, we asked if they could explain how their decision would change. Only 55\% of respondents who indicated some level of concern and had applied to a company partaking in related activities explicitly indicated that they were less likely to apply to roles at the particular companies. A few respondents argued that the moral responsibility lay with third parties such as managers (``\textit{I feel the onus is primarily on [other party]}'') and legislative authorities (``\textit{[I] am hoping that there is legistlation introduced to combat it}'').

Students commonly argued that even if they did not apply to the position, another employee would be hired to do the job: ``\textit{If I don't work for them, they'll hire an Indian for pennies anyways and put out infinitely worse stuff.}'' This comment not only refers to the offshoring of jobs as discussed previously, but also highlights the recent controversy surrounding H1B visas in the US. The majority of H1B visa holders come from India and China \cite{sherman2025h1b} and many have openly accused the visa scheme of depressing wages and stealing jobs from American workers \cite{gooding2025h1b} while tech companies (e.g., Microsoft) claim that the visas are essential to attracting top talent while maintaining the American tech industry’s competitive edge \cite{sadovi2025h1b,fishman2025h1b}. The views on foreign workers ranged from the more run-of-the-mill controversy (as highlighted above) to very strong: ``\textit{Every single time someone has ever proposed we bring manufacturing back and stop exploiting the rest of the world through imperialism, they get called fascists or neo-nazis or patriotic idiots ... Isolationism is the most moral and anti-imperialist policy, but it weakens you geopolitically and is unfavorable to most people.}''

\subsubsection{What is it about [Company] that makes you feel differently about [Concern]?}
If it was indicated that the knowledge about company activities would not change their decision, we asked them to explain their reasoning.

Approximately one fifth of respondents indicated that the prestige and salary offered by these corporations trumped their ethical concerns (``\textit{The brand/name recognition of (as well as all the opportunities that come with) these Big Tech companies makes it difficult to not want to seek opportunities there, especially early on in my career}'' and \textit{``The benefits of network and growth opportunities outweigh my personal contribution [to unethical actions]''}). An equal number indicated that while ethics did matter to them, they needed to do more research or were engaging in a more nuanced consideration of what their specific role would be. An equal number of respondents expressed that it was impossible to avoid companies engaging in the concerning action (``\textit{When a company gets large enough, it's inevitable they'll end up [partaking in the concern]}'') or used other moral argumentation to justify their choice.

Responses varied according to specific areas of concern. Those who were concerned about military use of their labour explained that their comfort at working for said companies greatly depended on what specifically their team was working on (e.g., ``\textit{If either directly developed arms I would feel differently}''). For those concerned about companies using exploitative working conditions, many felt that it was impossible to find work if they excluded companies engaged in exploitative practices. These responses differ from the responses provided by respondents concerned with privacy, many of whom indicated that the prestige, salary, and career opportunities matter more than the ethical concerns (e.g., ``\textit{The pay and benefits are miles ahead of non-FAANG companies. Also, just having the name [company] significantly increases the strength of a resume.}''). Likewise, these responses also differed from the responses of those concerned about mental health who argued that the fundamental moral responsibility fell upon the users, law-makers, or society (e.g., ``\textit{People need more self control}'' and ``\textit{This is more about individual control [than corporate responsibility]}'').

\subsubsection{What would it take to make you reconsider working for [Company]?}
If the respondent indicated that they were unsure if they would act on the knowledge presented to them about company activities, we asked them to describe what it would take for them to reconsider working for said companies.

Many respondents indicated that it depended on the specifics of role (e.g., ``\textit{providing [...] cloud services to the military would not be sufficient to stop me from applying, but it would be a factor in accepting a role and I would keep it in mind ...}'') and situation (e.g., ``\textit{The extent to which [Company] goes to lobby against climate change could change my views}''). These responses demonstrate that students believe that the actions may be unethical in theory but are willing to overlook them in practice, under specific circumstances. The responses also speak to the alienation of workers from their labour \cite{so2025cruel}. 
On the other hand, some respondents indicated that money/prestige/career were too important (\textit{``More money (and to an extent, prestige) means further compromised morals. Unfortunately''}), possibly reflecting the prestige bias observed in hiring practices \cite{chua2020you,biehl2025prestige}. This bias reinforces students' beliefs that to get ahead in what they perceive as difficult economic times, they have no choice but to compromise their ethical principles \cite{chua2020you}.

There were some demographic differences in the types of responses provided. Canadian respondents were much more likely to justify their choices by expressing desperation for a job (\textit{``I desperately [need] some sort of experience before I graduate,''}). US respondents were much more likely to indicate that the remuneration/career progression mattered more (i.e., less economic desperation) as reflected by: (\textit{``my application to [Company] is contingent on how much they are willing to buy my morals from me''}) or engage in moral disengagement and defeatism with regards to the matters which concerned them (\textit{``Honestly, it's too easy to find out that a company is not reducing their climate impact [...] Feels like it's impossible to individually make a positive climate impact when businesses are unable to be held accountable''}).

\section{Discussion}
 
\subsection{Apathy}
Many of the respondents were seemingly unconcerned with any of the listed issues. One possible reason for lack of concern is that the students surveyed tend to be insulated from the effects of the issues raised (compared to factory workers or those in the global south). The majority of the ethical issues examined do not directly impact Canada or the US (i.e., the first-order effects of militarism, exploitation of labour, climate change are likely not to initially occur in Canada and US to the same degree as other countries). In addition to geographic distance, demographic shields (those pursuing STEM degrees tend to have higher economic status \cite{thomas2021parental}) 
further protect them from the down-stream impacts of their work. 

Furthermore, the observed lack of concern across issues may also reflect differences in value alignment between respondents and the survey framing. The set of issues considered, while leveraged from past work, implicitly assumes that these topics represent salient ethical concerns in computing practice. Respondents may prioritize different ethical considerations that we did not capture: \textbf{low levels of concern does not necessarily indicate indifference to ethical questions, but may instead reflect a misalignment between the survey’s framing and students' ethical priorities}. In the appendix we discuss alignment between the concerns discussed in the survey and those expressed by respondents.

The free-text responses largely reflect dynamics previously observed in ethics education as studied by \citet{abdurrahman2023amazon}, namely how bids for ideological expansion (such as asking students to reflect on their decision to engage with unethical entities) can be rejected in a myriad of ways: to deflect and dismiss, affirm and move on, to reframe the conversation and most commonly seen in this study, to defer responsibility. 

Ethics education is, of course, not to blame for apathy nor is it the only lever available to tackle this apathy; structural constraints shape what any individual intervention can realistically achieve. Job-seeking students perceive entry-level opportunities outside of ethically contentious companies as limited, and face financial pressures which meaningfully narrow their possible choices. Complementary interventions worth exploring include (but are not limited to): collective organizing efforts among tech workers \cite{sum2025future}, and institutional support for students seeking roles in public interest technology, nonprofits, or government (akin to how loans for law graduates are forgiven for service to public institutions \cite{smith2023save}). While ethics education remains important to enabling student to act on their moral principles, its impact is likely to be greatest when students have real alternatives to choose from.

Taking a broader context, we note that most social activism (at the student and faculty level) has historically not come from those in STEM. For example, when professors at the University of Michigan were organizing a `work moratorium' in protest of the war in Vietnam, the majority of signatories were professors in the social sciences and humanities, ``most without the security of tenure'' \cite{tobinndteachin}. This has also been re-observed with recent pro-Palestine encampments, where most signatories of letters of support appear to be from the humanities (e.g., see \citet{brownfaculty2023ceasefire}), and most tenure denials stemming from (support for) activism appear to affect the humanities more \cite{ScottJPalestine}. Thus, apathy may be more indicative of a field-wide aversion to social causes.

\subsection{Effects of Curriculum on Student Concerns}

Universities typically position themselves not only as providers of technical training but as institutions that develop graduates into thoughtful, ethical contributors to society; a goal reflected in the mission statements of many CS programs as well as curricular guidelines \cite{kumar2024computer}. From this perspective, if students graduate without developing the ability to act upon and reason through the ethical dimensions of their professional lives, that represents a shortcoming of their education.

Prior work found required ethics courses in only 33\% of surveyed universities \cite{weichert2025have}. This was not the case with our sample. Examining the degree requirements of institutions from which we received multiple responses (covering approximately 83\% of respondents) showed that 56\% had a required standalone ethics course, while the remainder (except for a single non-ABET accredited institution) incorporated ethical education content through mandatory course modules. Thus, ethics instruction, in some form, was present virtually across the board.

\begin{figure*}[htb]
        \centering
        \includegraphics[width=0.7\textwidth]{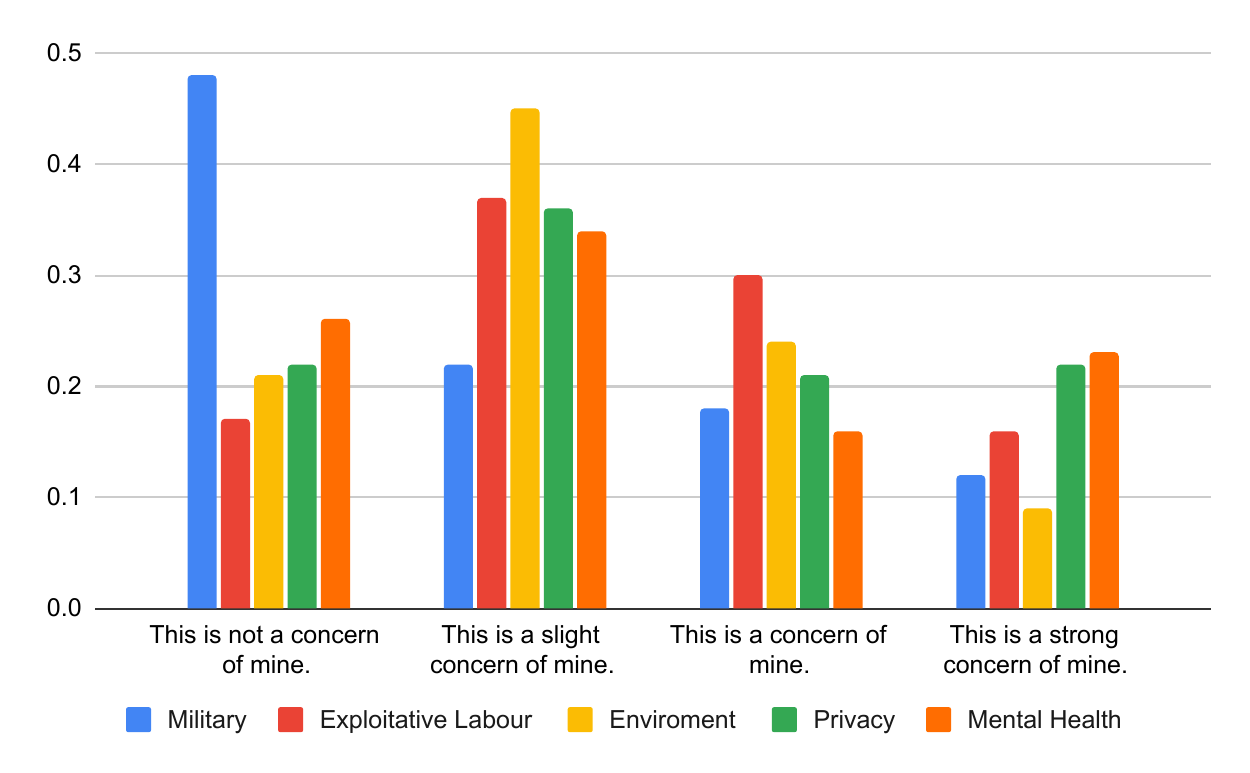}
        \caption{\label{fig:all_self_views} Decimal percentage of respondents (recently graduated or in their 4th year) who selected each level of concern for each issue collated in a single figure (individual views only).}
\end{figure*}

While ethics instruction is not solely responsible for students' ethical reasoning, we found that the variation across our sample in how concerned students are about the different ethical issues (Appendix Figure \ref{fig:all_self_views}) corresponds with what we know about the topics covered on typical CS ethics syllabi (though syllabi do not necessarily capture everything covered in a course). \citet{fiesler2020we} found that none of the syllabi they examined addressed military applications of technology, while just over half discussed privacy and surveillance. This coverage in syllabi could partially explain why concerns about labor used for military purposes ranked among the lowest of all issues here -- a finding that, at first glance, appears to contradict prior work suggesting that military applications of one's labor are at least as prominent a concern as privacy, surveillance, or inequality, in a study of the ethical concerns of (already employed) software engineers \cite{widder2023s}.

We looked at CS ethics syllabi from institutions with the six largest respondent groups (covering 60\% of respondents). These addressed both micro-ethics (professional conduct and standards) and macro-ethics (broader societal responsibility), and several used past industry scandals as case studies, or discussed how regulatory environments shape company behavior. Notably, only one syllabus explicitly framed ethical discussion in terms that positioned students as moral agents with responsibilities as future workers -- giving them a personal stake in the choices they will face professionally.

We agree with prior work \cite{sarder2022entering,raji2021you,brown2024teaching,thompson2025embracing} that explicitly preparing students to navigate ethical decisions in their working lives -- including decisions about where and for whom to work -- is a direction worth taking more seriously in CS ethics courses, as overall curricula do not appear to impart this skill-set. Rethinking tools, modes of thinking, and courses to support moral agency, and whether doing so shifts student behavior, is a question we hope future work will take up.

\subsection{Perceived Gaps in Ethical Engagement}
While there was a large sense of apathy, when exploring the difference between a respondent's self views and the perceived view of the field, we observed that, for all issues except exploitative working conditions, there was an over-estimation of the sense of apathy. For those most concerned about their labour being used for military purposes, there was a severe underestimation in the level of strong concern among others.

This large gap between respondents' self views and their perceived view of their field could have several explanations. One possible explanation is that the ethical questions raised may be viewed as political, and in STEM fields political advocacy can be seen as violating professional norms \cite{tormos2023pathways}. As such, students who have such concerns may feel uncomfortable expressing them publicly lest they risk their future careers: a student who assisted in this work requested not to be acknowledged fearing that working on a paper analyzing the intersection of the workplace, personal ethics and wider politics would compromise their chance at a successful career -- another manifestation of the desperation for employment felt by respondents.

Another possible explanation for the overestimation of apathy among fellow computer scientists could be a lack of observable actions aligned with ethical principles. If their peers, like most respondents, rationalize personal compromises, it may be difficult for individuals to recognize that others share their ethical concerns when those concerns are not acted upon. Furthermore, students and workers do not see ethical behaviour rewarded in CS (be it recognition from peers, honours from institutions, or material gains), thus placing ethical behaviour outside the norm of the ideal CS person \cite{darling2024not}. In fact, the persecution of ethical actors (e.g., whistle-blowers or protesters) 
may actually deter workers from reflecting too long on the implications of their day-to-day work. \citet{schiff2021linking} echo similar findings, noting that students feel a growing disconnect between their careers and their ethical tendencies: “\textit{Several interviewees also seemed to expect or hope that their social responsibility activities would fall into place eventually… academic success and career growth appear to be the clear priorities}”.

\subsection{Sub-group Differences}
Our survey highlighted some differences between respondents based on self-identified groups. Due to sample size issues, statistical testing for sub-group differences based on specific ethical issues were not robust. Corresponding percentages and p-values should be interpreted as exploratory signals rather than as confirmed, generalizable effects -- larger sample sizes are needed to confirm (or deny) the existence of any trends. For ethnicity, we limited our analysis to White and Asian due to sample size issues. While there was little difference between the groups on overall prioritization of concerns, for some responses, e.g., rate of application to Big Tech, the differences were significant. These differences could be caused by a wide variety of socio-cultural factors (e.g., immigration status/time, societal expectations, etc). Such a comparison is sensitive and must be done carefully to avoid falling prey to stereotypes\footnote{Such as the controversy which arose after a talk at NeurIPS framed Asian students as being less ethical/less likely to be taught ethics \cite{ha2024chinese}.}. The intention of this analysis was to highlight that there are possible differences among groups that may help inform curriculum development efforts, not to make generalizable claims about entire groups of people.

There were also differences observed between Male and non-Male respondents. Non-Male respondents valued location and ethical concerns more and were consistently less likely to state having no concern for the ethical issues surveyed (which held true for all except mental health). These differences, some of which are statistically significant, are in line with recent work highlighting the growing political differences between young men and women \cite{burn2024new}.

We also observed some significant differences between schools (e.g., rate of application to Big Tech). Concerns regarding the use of labour varied between schools and were highly dependent on the specific issue. These differences could have various explanations. For example, co-op programs at the Universities of Waterloo and Toronto may create a social pressure to pursue prestigious positions (i.e., ``Cali or Bust'') and could affect the way corporations interact with the schools (e.g., active recruitment and partnership). It may also be that the surrounding community, campus culture, and job market could affect student behaviour (e.g., living in a large city may result in a positive outlook on future opportunities compared to living in locales with limited local employment opportunities).

We observed the most pronounced differences between U.S. and Canadian respondents in the qualitative analysis. Specifically, Canadians were more likely to frame economic concerns in terms of financial insecurity, while American respondents more often highlighted the importance of financial reward (i.e., remuneration). Very strong descriptive support for the military was also more frequently expressed by American respondents. 

\subsection{Implications of Our Work}
Our survey reveals that while students have high levels of awareness regarding the possible ethical and societal issues they may encounter in their employment post-graduation, this awareness does not necessarily translate into effective ethical decision making, echoing similar findings in previous work \cite{sarder2022entering,raji2021you}. Students appear to struggle to apply or prioritize this knowledge in their own personal contexts; our findings indicate a heavy reliance on moral disengagement. This disconnect between awareness and ability to act underscores the importance of equipping them with tools to critically evaluate situations, recognize flawed reasoning, and confidently apply their understanding in practice. We believe that educators could help their students by presenting them more realistic ethical conundrums (such as those faced during the job search) earlier in their education, allowing them to consider possible (and more ethically consistent) responses before encountering them in real life.

For corporations: while many students appear unconcerned with the issues presented in this work, it is clear that most of those with concerns are still going to apply to companies whose actions they disagree with, despite their reservations. Given that it is unlikely companies will be able to determine which applicants feel this way (as they will likely not self-identify), this presents a risk to corporations who may unknowingly face ethical conflicts with a sizable portion of their workforce. It is also worth noting that respondents regularly underestimated the percentage of their peers who shared their ethical concerns, which could have implications for those organizing worker resistance actions.

\paragraph{Future Work}
Formalized adoption and expansion of this survey would increase the robustness of our results and enable more indepth analysis as well as longitudinal analysis.

\section{Conclusion}
In this work, we explored the decision-making processes of current or recently-graduated job-seeking CS students studying in Canada and the US with respect to various ethical issues. We found that ethical concerns ranked last among all of the possible considerations students could select, while remuneration ranked highest. Most students had applied to at least one of the Big Tech companies. The justifications provided for the decisions made during their job searches highlight a possible gap between ethics education and students' ability (or willingness) to apply the lessons to the decisions they face in their working lives. Respondents were very sensitive to external economic pressures, with some even viewing the search for employment with desperation, which resulted in moral compromises. Those compromises were often justified using a common set of rationalization strategies. These findings have important implications for educators, who should consider changes to curriculum to directly address the moral dilemmas that are being faced by new graduates and their tendency to rationalize their decisions. The findings also have implications for activists who could gain a better understanding of the outlook of new employees, and for corporations investing in their future workforce. We hope that this work will serve as a pilot study for a larger survey conducted by a professional or scholarly organization with the influence to reach a wider sample of respondents.

\section{Acknowledgments}
We would like to acknowledge the contributions of a student who supported this work but chose to remain anonymous out of concern that it could negatively affect their career due to its perceived political nature. We hope that future students will not feel compelled to make the same choice.

Mohamed Abdalla is supported through funding from University of Alberta and a CIFAR AI chair. Catherine Stinson and Alicia Cappello were supported by the Office of the Privacy Commissioner.

\bibliography{aaai2026}

\newpage
\section{Appendix}
\begin{table*}[h!]
\scalebox{0.8}{
\begin{tabular}{l|cccccccccccc}
\textbf{\% \textbackslash Respondent Type} & \textbf{All} & \textbf{\begin{tabular}[c]{@{}c@{}}4th Year \\ /Graduated\end{tabular}} & \textbf{\begin{tabular}[c]{@{}c@{}}Canada \\ (4th)\end{tabular}} & \textbf{\begin{tabular}[c]{@{}c@{}}US \\ (4th)\end{tabular}} & \textbf{\begin{tabular}[c]{@{}c@{}}Male\\ (4th)\end{tabular}} & \textbf{\begin{tabular}[c]{@{}c@{}}Non-male\\ (4th)\end{tabular}} & \textbf{\begin{tabular}[c]{@{}c@{}}White\\ (4th)\end{tabular}} & \textbf{\begin{tabular}[c]{@{}c@{}}Asian\\ (4th)\end{tabular}} & \textbf{\begin{tabular}[c]{@{}c@{}}UofT\\ (4th)\end{tabular}} & \textbf{\begin{tabular}[c]{@{}c@{}}Queen’s\\ (4th)\end{tabular}} & \textbf{\begin{tabular}[c]{@{}c@{}}Waterloo\\ (4th)\end{tabular}} & \textbf{\begin{tabular}[c]{@{}c@{}}1-3rd\\ Year\end{tabular}} \\ \hline
\textbf{Denom (n)}& 129& 90& 61 & 29  & 55& 35  & 40 & 40 & 19& 15 & 8   & 39   \\
\textbf{At least 1 BT:} & 65.12   & 70& 67.21& 75.86& 72.73& 65.71 & 57.5  & 82.5  & 78.95& 40 & 87.5& 53.85\\
\textbf{Alphabet}  & 47.29   & 53.33  & 44.26& 72.41& 50.91& 57.14 & 42.5  & 62.5  & 68.42& 26.67& 50  & 33.33\\
\textbf{Amazon}    & 51.16   & 57.78  & 52.46& 68.97& 58.18& 57.14 & 42.5  & 72.5  & 52.63& 40 & 87.5& 35.9 \\
\textbf{Apple}& 41.09   & 46.67  & 39.34& 62.07& 40& 57.14 & 32.5  & 60 & 36.84& 26.67& 75  & 28.21\\
\textbf{Meta} & 39.53   & 42.22  & 31.15& 65.52& 40& 45.71 & 30 & 55 & 31.58& 26.67& 62.5& 33.33\\
\textbf{Microsoft} & 51.16   & 56.67  & 50.82& 68.97& 52.73& 62.86 & 42.5  & 70 & 52.63& 33.33& 62.5& 38.46
\end{tabular}
}
\caption{\label{tab:application_percentage2} Percent of respondents who applied to any (and each) of the listed companies split by selected sub-group. ``4th'' indicates that only respondents in their fourth year or those who recently graduated are considered in the column. BT = Big Tech.}
\end{table*}

\begin{table*}[h!]
\scalebox{0.8}{
\begin{tabular}{l|cccccccccccc}
\textbf{Respondent Type} & \textbf{All} & \textbf{\begin{tabular}[c]{@{}c@{}}4th Year \\ /Graduated\end{tabular}} & \textbf{\begin{tabular}[c]{@{}c@{}}Canada \\ (4th)\end{tabular}} & \textbf{\begin{tabular}[c]{@{}c@{}}US \\ (4th)\end{tabular}} & \textbf{\begin{tabular}[c]{@{}c@{}}Male\\ (4th)\end{tabular}} & \textbf{\begin{tabular}[c]{@{}c@{}}Non-male\\ (4th)\end{tabular}} & \textbf{\begin{tabular}[c]{@{}c@{}}White\\ (4th)\end{tabular}} & \textbf{\begin{tabular}[c]{@{}c@{}}Asian\\ (4th)\end{tabular}} & \textbf{\begin{tabular}[c]{@{}c@{}}UofT\\ (4th)\end{tabular}} & \textbf{\begin{tabular}[c]{@{}c@{}}Queen’s\\ (4th)\end{tabular}} & \textbf{\begin{tabular}[c]{@{}c@{}}Waterloo\\ (4th)\end{tabular}} & \textbf{\begin{tabular}[c]{@{}c@{}}1-3rd\\ Year\end{tabular}} \\ \hline
\textbf{Remuneration} & 2.31    & 2.47& 2.56 & 2.29  & 2.35   & 2.68  & 2.49    & 2.4& 2.53   & 2.87 & 2.38  & 1.92 \\
\textbf{Location}& 3.47    & 3.46& 3.46 & 3.46  & 3.76   & 2.97  & 3.46    & 3.4& 3.63   & 4    & 3.25  & 3.5  \\
\textbf{Work Life Balance} & 3.38    & 3.3 & 3.3  & 3.32  & 3.31   & 3.29  & 3.33    & 3.2& 3.89   & 2.73 & 3.62  & 3.55 \\
\textbf{Workplace Culture} & 4.09    & 4.1 & 4.21 & 3.86  & 4.04   & 4.21  & 4.08    & 4.12    & 4.05   & 3.93 & 3.75  & 4.05 \\
\textbf{Ethical Concerns}  & 5.01    & 5.09& 5.07 & 5.14  & 5.29   & 4.76  & 5.1& 5.38    & 5.11   & 4.73 & 5.38  & 4.82 \\
\textbf{Prof. Dev. Opps}   & 3.26    & 3.22& 3.11 & 3.46  & 3.04   & 3.53  & 3.36    & 3.15    & 3.16   & 3.53 & 2.62  & 3.34 \\
\textbf{Other}   & 6.49    & 6.35& 6.3  & 6.46  & 6.22   & 6.56  & 6.18    & 6.35    & 5.63   & 6.2  & 7& 6.82
\end{tabular}
}
\caption{\label{tab:consideration_avgs2} Average rank of job considerations for all respondents with analyses for different sub-groups (gender, ethnicity, and university).}
\end{table*}

\subsection{Survey Details}
\label{app:surveymethods}
 The survey is composed of three sections. The first section dealt with demographic information, companies applied to, and ranking a predetermined list of priorities from 1 to 7, with 1 being the most important consideration and 7 being the least important consideration. Respondents were also able to add a priority not found on the predetermined list.

The second section of the survey asked respondents if they had any concerns about their labour being used: 1) for military purposes, 2) by a company engaged in exploitative working conditions (e.g., forced labour or wage theft), 3) to negatively impact the environment, 4) to invade the privacy of others, or 5) to negatively impact the mental health of others. This selection of ethical issues was informed by past work \cite{widder2023open}. For each of these concerns, respondents could choose from 4 different options ranging from indifference to strong concern: 1) "This is not a concern of mine," 2) "This is a slight concern of mine. I would not apply to departments/teams which are known to \textbf{directly participate in \textit{[area of concern]}.} I would still apply to other departments/teams at such a company," 3) ``This is a concern of mine. I would not apply to departments/teams which \textbf{may participate in \textit{[area of concern]}.} I would still apply to other departments/teams at such a company," and 4) "This is a strong concern of mine. I would not apply to any company which is known to participate in \textit{[area of concern]}". For each issue, respondents were also asked to select a statement which they believe best described the views of the field (i.e., most job-seeking CS graduates) and were offered a text box to explain their selections (if desired).

The last section of the survey was dynamic, presenting each respondent with different questions by comparing their responses about companies applied to (first section) with their stated levels of concern (second section). If a conflict was detected (e.g., applying to work for Alphabet or Amazon while having selected that labour being used for military purposes was a strong concern --- given the well documented military contracts awarded to these companies \cite{gonzalez2024big}) --- respondents were asked if they were aware of the conflict (possible responses: 1) ``Yes, I was aware of all of them,'' 2) ``Yes, I was aware of some of them,'' or 3) ``No, this is new to me''). Next, the survey asked if this knowledge changed their opinion of applying to the relevant company (possible responses: 1) ``Yes,'' 2) ``No,'' or 3) ``I'd have to think about it''). If respondents selected "Yes," the survey asked them to expand on how this knowledge changed their opinion of applying to the relevant company. If respondents selected "No," the survey asked them what was about the company or role that makes them feel differently. If the respondents selected "I'd have to think more about it," the survey asked them what it would take for them to reconsider working for the relevant company. 

To mitigate participation bias (also known as non-response bias), which occurs when the characteristics of the survey sample differ from the characteristics of the overall population \cite{sedgwick2014non}, survey participants were told that the purpose of the survey was to assist CS educators understand student preferences and views related to the job search. To avoid order effects bias, which occurs when a respondent's answers are affected by the order of the questions \cite{perreault1975controlling}, the questions were structured such that more innocuous questions (e.g., `What companies have you applied to?') were asked before questions more likely to illicit a defensive response (e.g., `What are your views of your labour being used for military purposes?'). To minimize framing effects --- changes in findings due to phrasing of questions --- we iteratively worked to improve the phrasing by beta-testing our survey among colleagues and personal connections several times. To enable replication, our full survey protocol is included as supplemental material.

\subsection{All rankings}

\subsection{Percentage of Respondents agreeing to specific statements}

\begin{table*}[h!]
\centering
\begin{tabular}{l|cccccccccccc}
\textbf{Respondent Type} & \textbf{All} & \textbf{\begin{tabular}[c]{@{}c@{}}4th Year \\ /Graduated\end{tabular}} & \textbf{\begin{tabular}[c]{@{}c@{}}Canada \\ (4th)\end{tabular}} & \textbf{\begin{tabular}[c]{@{}c@{}}US \\ (4th)\end{tabular}} & \textbf{\begin{tabular}[c]{@{}c@{}}Male\\ (4th)\end{tabular}} & \textbf{\begin{tabular}[c]{@{}c@{}}Non-male\\ (4th)\end{tabular}}  \\ \hline
\textbf{No Concern} & 46.90   & 47.5& 46.15& 50.00 & 58.00  & 30.00 & \\
\textbf{Slight Concern}  & 23.01   & 22.5& 23.08& 21.43 & 22.00  & 23.33 \\
\textbf{Concern}    & 17.70   & 17.5& 17.31& 17.86 & 14.00  & 23.33 & \\
\textbf{Strong Concern}  & 12.39   & 12.5& 13.46& 10.71 & 6.00   & 23.33 \\
\textbf{Denom} & 114& 80  & 52   & 28    & 50& 30  \\ \\  
\textbf{Respondent Type} & \textbf{\begin{tabular}[c]{@{}c@{}}White\\ (4th)\end{tabular}} & \textbf{\begin{tabular}[c]{@{}c@{}}Asian\\ (4th)\end{tabular}} & \textbf{\begin{tabular}[c]{@{}c@{}}UofT\\ (4th)\end{tabular}} & \textbf{\begin{tabular}[c]{@{}c@{}}Queen’s\\ (4th)\end{tabular}} & \textbf{\begin{tabular}[c]{@{}c@{}}Waterloo\\ (4th)\end{tabular}} & \textbf{\begin{tabular}[c]{@{}c@{}}1-3rd\\ Year\end{tabular}} \\ \hline
\textbf{No Concern} & 52.94   & 50.00   & 37.50  & 53.85& 57.14 & 44.12\\
\textbf{Slight Concern}  & 26.47   & 22.22   & 31.25  & 15.38& 28.57 & 23.53\\
\textbf{Concern}    & 5.88    & 16.67   & 18.75  & 15.38& 0.00  & 20.59\\
\textbf{Strong Concern}  & 14.71   & 11.11   & 12.5   & 15.38& 14.29 & 11.76\\
\textbf{Denom} & 34 & 36 & 16& 13   & 7& 34
\end{tabular}
    \captionof{table}{Percentage of respondents agreeing with specific statements regarding the use of their labour for military purposes, split by various sub-groups.}
    \label{tab:percentage_military_views_nopair}
\end{table*}

\begin{table*}[h!]
\centering
    \begin{tabular}{l|cccccccccccc}
\textbf{Respondent Type} & \textbf{All} & \textbf{\begin{tabular}[c]{@{}c@{}}4th Year \\ /Graduated\end{tabular}} & \textbf{\begin{tabular}[c]{@{}c@{}}Canada \\ (4th)\end{tabular}} & \textbf{\begin{tabular}[c]{@{}c@{}}US \\ (4th)\end{tabular}} & \textbf{\begin{tabular}[c]{@{}c@{}}Male\\ (4th)\end{tabular}} & \textbf{\begin{tabular}[c]{@{}c@{}}Non-male\\ (4th)\end{tabular}}  \\ \hline
\textbf{No Concern} & 58.05   & 17.11    & 18.00& 15.38 & 20.41  & 11.11 \\
\textbf{Slight Concern}  & 18.54   & 36.84    & 34.00& 42.31 & 36.73  & 37.04 \\
\textbf{Concern}    & 14.63   & 30.26    & 36.00& 19.23 & 30.61  & 29.63 \\
\textbf{Strong Concern}  & 8.78    & 15.79    & 12.00& 23.08 & 12.24  & 22.22  \\
\textbf{Denom} & 106 & 76 & 50 & 26 & 49 & 27 \\ \\ 
\textbf{Respondent Type} & \textbf{\begin{tabular}[c]{@{}c@{}}White\\ (4th)\end{tabular}} & \textbf{\begin{tabular}[c]{@{}c@{}}Asian\\ (4th)\end{tabular}} & \textbf{\begin{tabular}[c]{@{}c@{}}UofT\\ (4th)\end{tabular}} & \textbf{\begin{tabular}[c]{@{}c@{}}Queen’s\\ (4th)\end{tabular}} & \textbf{\begin{tabular}[c]{@{}c@{}}Waterloo\\ (4th)\end{tabular}} & \textbf{\begin{tabular}[c]{@{}c@{}}1-3rd\\ Year\end{tabular}} \\ \hline
\textbf{No Concern} & 17.65   & 15.62   & 13.33  & 38.46& 0.0   & 23.33\\
\textbf{Slight Concern}  & 35.29   & 40.62   & 40.00  & 30.77& 33.33 & 33.33\\
\textbf{Concern}    & 32.35   & 28.12   & 40.00  & 30.77& 16.67 & 23.33\\
\textbf{Strong Concern}  & 14.71   & 15.62   & 6.67   & 0.0  & 50.00 & 20   \\
\textbf{Denom} & 34 & 32 & 15 & 13 & 6 & 30 
\end{tabular}
    \captionof{table}{Percentage of respondents who agreed with specific statements regarding the use of their labour by companies known to engage in exploitative working conditions, split by various sub-groups.}
    \label{tab:percentage_exploit_views_nopair}
\end{table*}

\begin{table*}[h!]
\centering
\begin{tabular}{l|cccccccccccc}
\textbf{Respondent Type} & \textbf{All} & \textbf{\begin{tabular}[c]{@{}c@{}}4th Year \\ /Graduated\end{tabular}} & \textbf{\begin{tabular}[c]{@{}c@{}}Canada \\ (4th)\end{tabular}} & \textbf{\begin{tabular}[c]{@{}c@{}}US \\ (4th)\end{tabular}} & \textbf{\begin{tabular}[c]{@{}c@{}}Male\\ (4th)\end{tabular}} & \textbf{\begin{tabular}[c]{@{}c@{}}Non-male\\ (4th)\end{tabular}}  \\ \hline
\textbf{No Concern} & 24.76   & 21.33    & 20.00& 24.00 & 24.49  & 15.38 \\
\textbf{Slight Concern}  & 41.90   & 45.33    & 42.00& 52.00 & 42.86  & 50.00 \\
\textbf{Concern}    & 23.81   & 24.00    & 26.00& 20.00 & 22.45  & 26.92 \\
\textbf{Strong Concern}  & 9.52    & 9.33& 12.00& 4.00  & 10.20  & 7.69  \\
\textbf{Denom} & 105 & 75 & 50 & 25 & 49 & 26  \\ \\ 
\textbf{Respondent Type} & \textbf{\begin{tabular}[c]{@{}c@{}}White\\ (4th)\end{tabular}} & \textbf{\begin{tabular}[c]{@{}c@{}}Asian\\ (4th)\end{tabular}} & \textbf{\begin{tabular}[c]{@{}c@{}}UofT\\ (4th)\end{tabular}} & \textbf{\begin{tabular}[c]{@{}c@{}}Queen’s\\ (4th)\end{tabular}} & \textbf{\begin{tabular}[c]{@{}c@{}}Waterloo\\ (4th)\end{tabular}} & \textbf{\begin{tabular}[c]{@{}c@{}}1-3rd\\ Year\end{tabular}} \\ \hline
\textbf{No Concern} & 15.15   & 28.12   & 20.00  & 23.08& 16.67 & 23.33\\
\textbf{Slight Concern}  & 45.45   & 43.75   & 33.33  & 46.15& 33.33 & 33.33\\
\textbf{Concern}    & 24.24   & 21.88   & 26.67  & 23.08& 33.33 & 23.33\\
\textbf{Strong Concern}  & 15.15   & 6.25    & 20.00  & 7.69 & 16.67 & 20.00\\
\textbf{Denom} & 33 & 32 & 15 & 13 & 6 & 30 
\end{tabular}
    \captionof{table}{Percentage of respondents agreeing with specific statements regarding the use of their labour by companies known to negatively impact the environment.}
    \label{tab:enviorment_views}
\end{table*}

\begin{table*}[h!]
\centering
\begin{tabular}{l|cccccccccccc}
\textbf{Respondent Type} & \textbf{All} & \textbf{\begin{tabular}[c]{@{}c@{}}4th Year \\ /Graduated\end{tabular}} & \textbf{\begin{tabular}[c]{@{}c@{}}Canada \\ (4th)\end{tabular}} & \textbf{\begin{tabular}[c]{@{}c@{}}US \\ (4th)\end{tabular}} & \textbf{\begin{tabular}[c]{@{}c@{}}Male\\ (4th)\end{tabular}} & \textbf{\begin{tabular}[c]{@{}c@{}}Non-male\\ (4th)\end{tabular}}  \\ \hline
\textbf{No Concern} & 22.33   & 21.92    & 22.45& 20.83 & 23.40  & 19.23  \\
\textbf{Slight Concern}  & 33.98   & 35.62    & 36.73& 33.33 & 36.17  & 34.62  \\
\textbf{Concern}    & 20.39   & 20.55    & 22.45& 16.67 & 19.15  & 23.08  \\
\textbf{Strong Concern}  & 23.30   & 21.92    & 18.37& 29.17 & 21.28  & 23.08 \\
\textbf{Denom} & 103 & 73 & 49 & 24 & 47 & 26  \\ \\ 
\textbf{Respondent Type} & \textbf{\begin{tabular}[c]{@{}c@{}}White\\ (4th)\end{tabular}} & \textbf{\begin{tabular}[c]{@{}c@{}}Asian\\ (4th)\end{tabular}} & \textbf{\begin{tabular}[c]{@{}c@{}}UofT\\ (4th)\end{tabular}} & \textbf{\begin{tabular}[c]{@{}c@{}}Queen’s\\ (4th)\end{tabular}} & \textbf{\begin{tabular}[c]{@{}c@{}}Waterloo\\ (4th)\end{tabular}} & \textbf{\begin{tabular}[c]{@{}c@{}}1-3rd\\ Year\end{tabular}} \\ \hline
\textbf{No Concern} & 18.75   & 29.03   & 13.33  & 25.00& 50.00 & 23.33\\
\textbf{Slight Concern}  & 31.25   & 45.16   & 26.67  & 41.67& 33.33 & 30.00\\
\textbf{Concern}    & 21.88   & 9.68    & 26.67  & 25.00& 0.0   & 20.00\\
\textbf{Strong Concern}  & 28.12   & 16.13   & 33.33  & 8.33 & 16.67 & 26.67\\
\textbf{Denom} & 32 & 31 & 15 & 12 & 6 & 30  
\end{tabular}
    \captionof{table}{Percentage of respondents who agreed with specific statements regarding the use of their labour by companies known to invade the privacy of others.}
    \label{tab:privacy_views}
\end{table*}

\begin{table*}[h!]
\centering
    \begin{tabular}{l|cccccccccccc}
\textbf{Respondent Type} & \textbf{All} & \textbf{\begin{tabular}[c]{@{}c@{}}4th Year \\ /Graduated\end{tabular}} & \textbf{\begin{tabular}[c]{@{}c@{}}Canada \\ (4th)\end{tabular}} & \textbf{\begin{tabular}[c]{@{}c@{}}US \\ (4th)\end{tabular}} & \textbf{\begin{tabular}[c]{@{}c@{}}Male\\ (4th)\end{tabular}} & \textbf{\begin{tabular}[c]{@{}c@{}}Non-male\\ (4th)\end{tabular}} \\ \hline
\textbf{No Concern} & 26.00   & 26.03    & 30.61& 16.67 & 23.4   & 30.77 \\
\textbf{Slight Concern}  & 32.00   & 34.25    & 26.53& 50.00 & 36.17  & 30.77 \\
\textbf{Concern}    & 20.00   & 16.44    & 16.33& 16.67 & 17.02  & 15.38 \\
\textbf{Strong Concern}  & 23.00   & 23.29    & 26.53& 16.67 & 23.40  & 23.08 \\
\textbf{Denom} & 100 & 73   & 49    & 24 & 47 & 26 \\ \\
\textbf{Respondent Type} &  \textbf{\begin{tabular}[c]{@{}c@{}}White\\ (4th)\end{tabular}} & \textbf{\begin{tabular}[c]{@{}c@{}}Asian\\ (4th)\end{tabular}} & \textbf{\begin{tabular}[c]{@{}c@{}}UofT\\ (4th)\end{tabular}} & \textbf{\begin{tabular}[c]{@{}c@{}}Queen’s\\ (4th)\end{tabular}} & \textbf{\begin{tabular}[c]{@{}c@{}}Waterloo\\ (4th)\end{tabular}} & \textbf{\begin{tabular}[c]{@{}c@{}}1-3rd\\ Year\end{tabular}} \\ \hline
\textbf{No Concern} & 34.38   & 19.3    & 20.00  & 46.15& 33.33 & 25.00\\
\textbf{Slight Concern}  & 21.88   & 45.16   & 20.00  & 23.08& 33.33 & 25.00\\
\textbf{Concern}    & 18.75   & 12.9    & 20.00  & 15.38& 0.0   & 28.57\\
\textbf{Strong Concern}  & 25.00   & 22.58   & 40.00  & 15.38& 33.33 & 21.43\\
\textbf{Denom} & 32 & 31 & 15 & 13 & 6 & 28
\end{tabular}
\captionof{table}{Percentage of respondents agreeing with specific statements regarding the use of their labour by companies known to negatively impact the mental health of others.}
    \label{tab:mental_views}
\end{table*}

\newpage
\section{Awareness of Ethical Issues}
Table \ref{tab:app_knowledge} presents the awareness of respondents (in their 4th year and recently graduated) to the ethical issues of companies they are applying to for moral issues they indicated being concerned about.

\begin{table*}[h!]
\centering
\begin{tabular}{l|cccc}
\textbf{}           & \multicolumn{1}{l}{\textbf{Yes, fully aware}} & \multicolumn{1}{l}{\textbf{Yes, partly Aware}} & \multicolumn{1}{l}{\textbf{No, this is new to me}} & \multicolumn{1}{l}{\textbf{\% with some level of awareness}} \\ \hline
Military & 3 & 3 & 6 & 50\% \\
Exploitative Labour & 3 & 9 & 5 & 71\% \\
Environment & 4 & 2 & 8 & 43\% \\
Privacy & 4 & 7 & 6 & 65\% \\
Mental Health & 8 & 2 & 2 & 83\%
\end{tabular}
\caption{\label{tab:app_knowledge}} The awareness of respondents (in their 4th year and recently graduated) to the ethical issues of companies they are applying to for moral issues they indicated being concerned about.
\end{table*}

\section{Alignment of Concerns}
\label{sec:alignment_appendix}
\begin{figure*}[h!]
    \setlength{\intextsep}{1pt}
    \setlength{\belowcaptionskip}{0pt}
    \centering
        \includegraphics[width=0.4\textwidth]{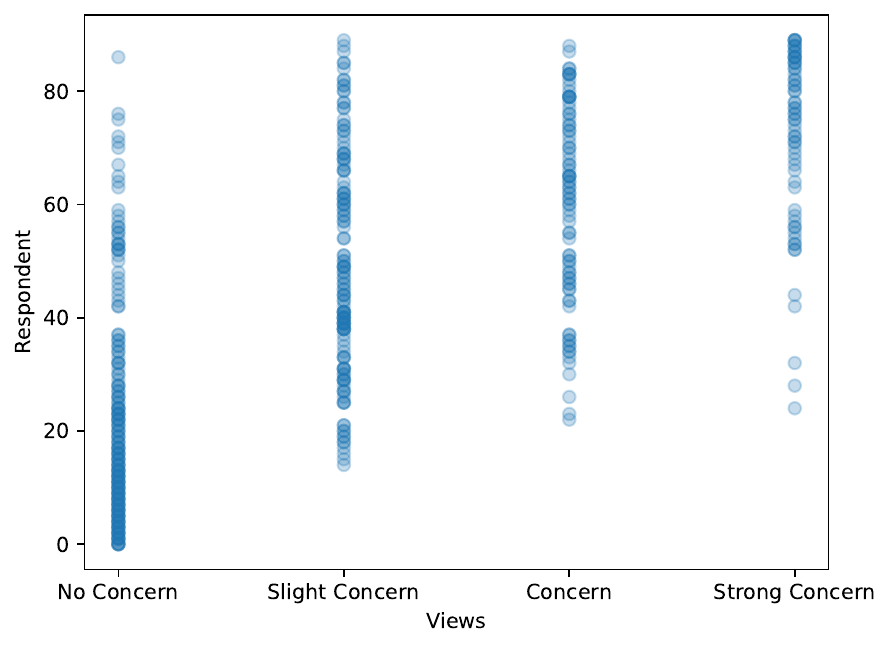}
    \caption{Self views for all respondents (recently graduated or in their 4th year) for all concerns. Each row represents the responses for a single respondent. As there are 5 areas of concern, each row will have 5 dots. Dots are plotted with an alpha of 0.2 such that we can tell where overlapping responses may be (e.g., a single response of `No Concern' will be very light whereas all 5 responses of `No Concern' will result in a dark blue circle. \label{fig:individual_views}}
\end{figure*}

Exploring the responses, we find that most respondents have large variability in their level of concern across ethical issues (though there are a large number of students seemingly unconcerned with any of the listed issues), as shown in Figure \ref{fig:individual_views}. Being very strongly concerned about one issue does not necessarily translate into a strong concern for the other issues. 
This may reflect differing ethical alignment with the set of issues covered. For example, military and environment might be seen as partisan issues, which survey respondents have divided views on, while mental health and privacy (of one's own citizens) are more bipartisan. More research covering a broader range of ethical issues would be needed to better understand how a student's alignment with particular ethical issues interacts with their  reasoning about them.

\clearpage
\section{Full Canadian Survey}
\label{sec:appendix_fullsurvery}
See next page.
\includepdf[pages=-, width=\textwidth, frame=true]{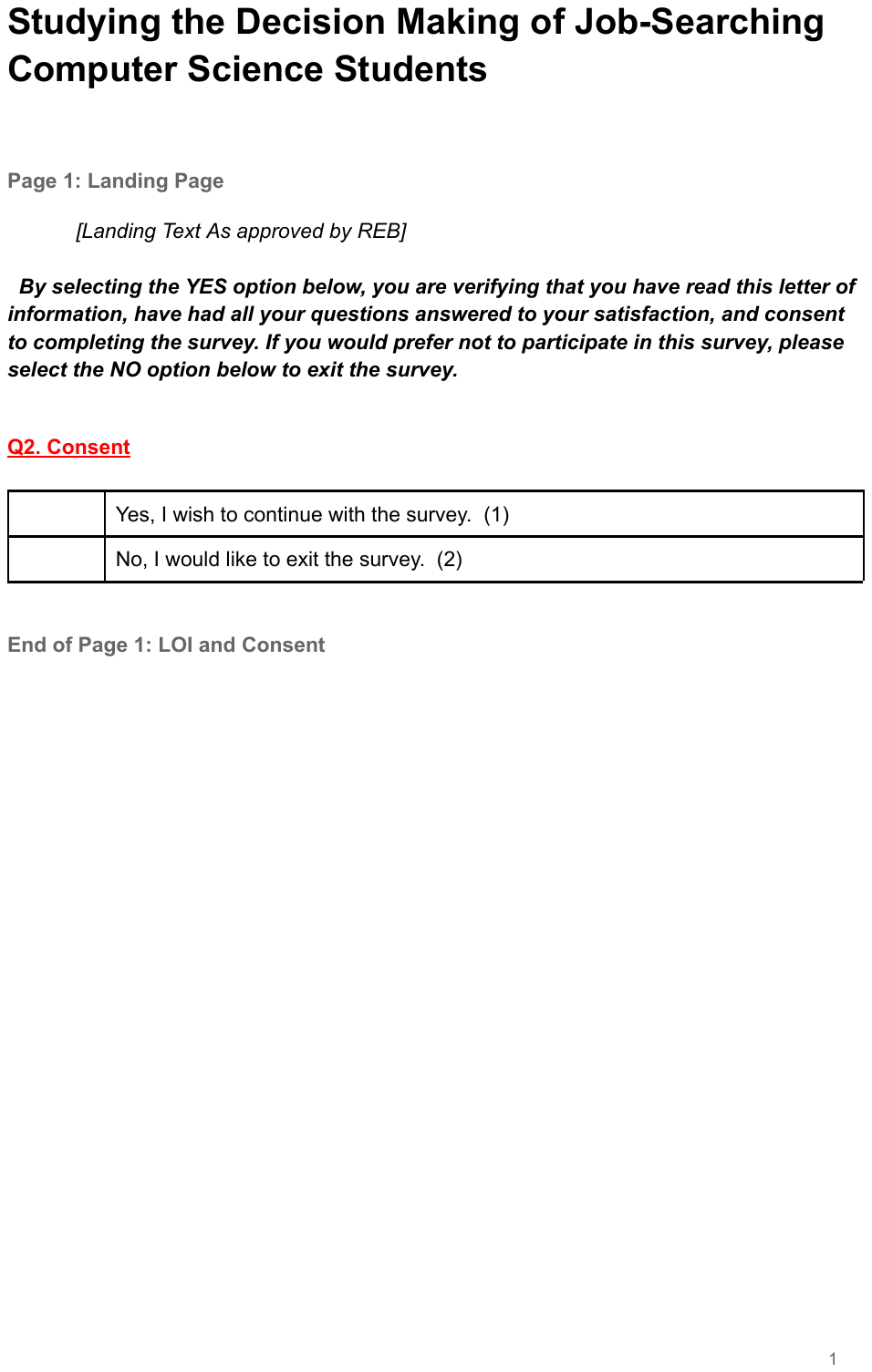}

\label{app:survey}

\end{document}